\documentclass[useAMS,usenatbib]{mn2e}

\usepackage{graphicx}
\usepackage{subfigure}

\title[On the  various origins  of close-in extrasolar  planets]{On the
various origins of close-in extrasolar planets}

\author[Marchi et al.]{S. Marchi$^{1}$\thanks{E-mail:
simone.marchi@unipd.it}, S. Ortolani$^{1}$
M. Nagasawa$^{2}$ and S. Ida$^{2}$\\
$^{1}$Department of Astronomy, Padova University, 34122 Padova, Italy\\
$^{2}$Tokyo Institute of Technology, Tokyo 152-8550, Japan}
\begin{document}

\date{Accepted ...}

\pagerange{\pageref{firstpage}--\pageref{lastpage}} \pubyear{2008}

\maketitle

\label{firstpage}

\begin{abstract}

%The  discovery  of  the  first  extrasolar planets  (EPs)  orbiting  a
%solar-like star  opened a new  exciting phase in our  understanding of
%planetary formation and evolution.  Soon after the first discovery, 
The extrasolar planets (EPs) so far detected are very different to the
planets  in our  own Solar  System.   Many of  them have  Jupiter-like
masses  and close-in  orbits (the  so-called hot  planets,  HPs), with
orbital periods of  only a few days.  In this paper,  we present a new
statistical analysis  of the observed  EPs, focusing on the  origin of
the HPs.
%For a comparison, giant gaseous  planets of our Solar
%System  formed beyond  some 3-4~AU  from the  Sun (namely,  beyond the
%so-called snow-line  of the protoplanetary disk), at  a distance close
%to  their present  semimajor axis.   
Among the  several HP  formation mechanisms proposed  so far,  the two
main formation  mechanisms are  type~II migration and  scattering.  In
both  cases,  planets  form  beyond  the so-called  snow-line  of  the
protoplanetary disk  and then migrate  inward due to  angular momentum
and  energy  exchange with  either  the  protoplanetary  disk or  with
companion planets.   Although theoretical  studies produce a  range of
observed features, no firm correspondence between the observed EPs and
models  has  yet been  established.   In  our  analysis, by  means  of
principal  component analysis  and hierarchical  cluster  analysis, we
find convincing  indications for  the existence of  two types  of HPs,
whose parameters reflect physical  mechanisms of type~II migration and
scattering.

\end{abstract}


\section{Taxonomy of hot planets}

The present EPs database consists  of a rather heterogeneous sample of
planets,  showing  great  variety  in  all  the  measured  quantities.
Statistical analysis  provides a necessary means  to find correlations
among various physical parameters  involved in planetary formation and
evolution.  Nevertheless, we caution that statistical analysis may not
be able  to disclose  important relationships due  to the  complex and
-mostly unknown-  interplay of the  involved parameters.  In  order to
overcome such problems, we  performed a global statistical analysis of
the  EPs, using  a novel  approach.  The  underlying idea  is  to find
groups of {\it similar} EPs  and to distinguish different EP groups on
the  basis of their  {\it diversity}.   In this  work, the  concept of
similarity  and  diversity among  EPs  is  quantified  by means  of  a
distance measure in the  multifold space of physical parameters.  This
goal has  been achieved with  the aid of principal  component analysis
and  hierarchical  cluster  analysis.   In  the  present  analysis  we
followed  the   procedure  used   by  \cite{mar07},  and   updated  in
\cite{mar08}.    The   database\footnote{Taken   from  The   Exoplanet
Enclycopaedia at  http://exoplanet.eu/} used in this  paper is updated
to  July 9$^{th}$,  2008.  We  restricted  our analysis  to those  EPs
having measurements for five {\it input variables}, that is: planetary
mass ($M_p$),  semimajor axis ($a$), eccentricity  ($e$), stellar mass
($M_s$) and metallicity ([Fe/H]).  Of 308 EPs (including Jupiter), 252
were finally selected  for the analysis.  The purpose  of our analysis
is to  identify planets which are  similar with respect  to the 5-fold
space of the input variables.  As  a result, 6 robust EP clusters have
been  identified.  Before  focusing  on HPs,  we  briefly outline  the
general nature of the clusters.\\
Cluster C1  is characterized by  sub- to jovian-like  $M_p$, $a<1$~AU,
and super-solar  [Fe/H].  Cluster  C2 has sub-jovian  $M_p$, $a<1$~AU,
sub-solar $M_s$  and sub-solar [Fe/H].   Both clusters C1 and  C2 also
have low mean eccentricity. Cluster  C3 is the least populated cluster
(14 EPs) and probably has  no strong significance except the fact that
it contains  many peculiar  EPs that for  different reasons  have been
rejected by the other clusters.   They are mostly jovian mass planets,
having a remarkably super-solar  [Fe/H].  Cluster C4 has mostly jovian
mass planets, a relatively large $a$, orbiting solar mass stars with a
widespread  [Fe/H]  distribution  characterized  by  sub-solar  values
(Jupiter belongs  to this cluster). Cluster C5  has super-jovian mass,
$a>1$~AU, super-solar $M_s$ and  sub-solar [Fe/H].  The same holds for
cluster C6, except  for its super-solar [Fe/H] and  its higher average
eccentricity.   All the  input  variables have  an  important role  in
defining the clusters, in particular $a$ and [Fe/H].
%  (histograms of
%these  variables   are  reported   at  the  end   of  the   draft,  in
%fig. \ref{allep}).
Beyond the general traits of the solutions, which will be described in
more detail  elsewhere, we focus in  this paper on the  HPs defined as
those  having  a period  less  than 12  days  (see  later for  further
comments on this selection).   According to the adopted definition, 69
HPs are present in the database.   The main results of the taxonomy is
that HPs have  been split into two main  groups, belonging to clusters
C1 and  C2.  In  addition, there  are a few  outliers placed  in other
clusters.
%
%This is the case  of HPs provided
%that the  taxonomy groups them into  two main clusters  (namely C1 and
%C2).  However, it  must be clear that these  clusters also contain non
%HPs.\\
%

\subsection{Two types of HPs}

\begin{figure*}
\centering
\subfigure{\includegraphics[width=0.4\textwidth,angle=-90]{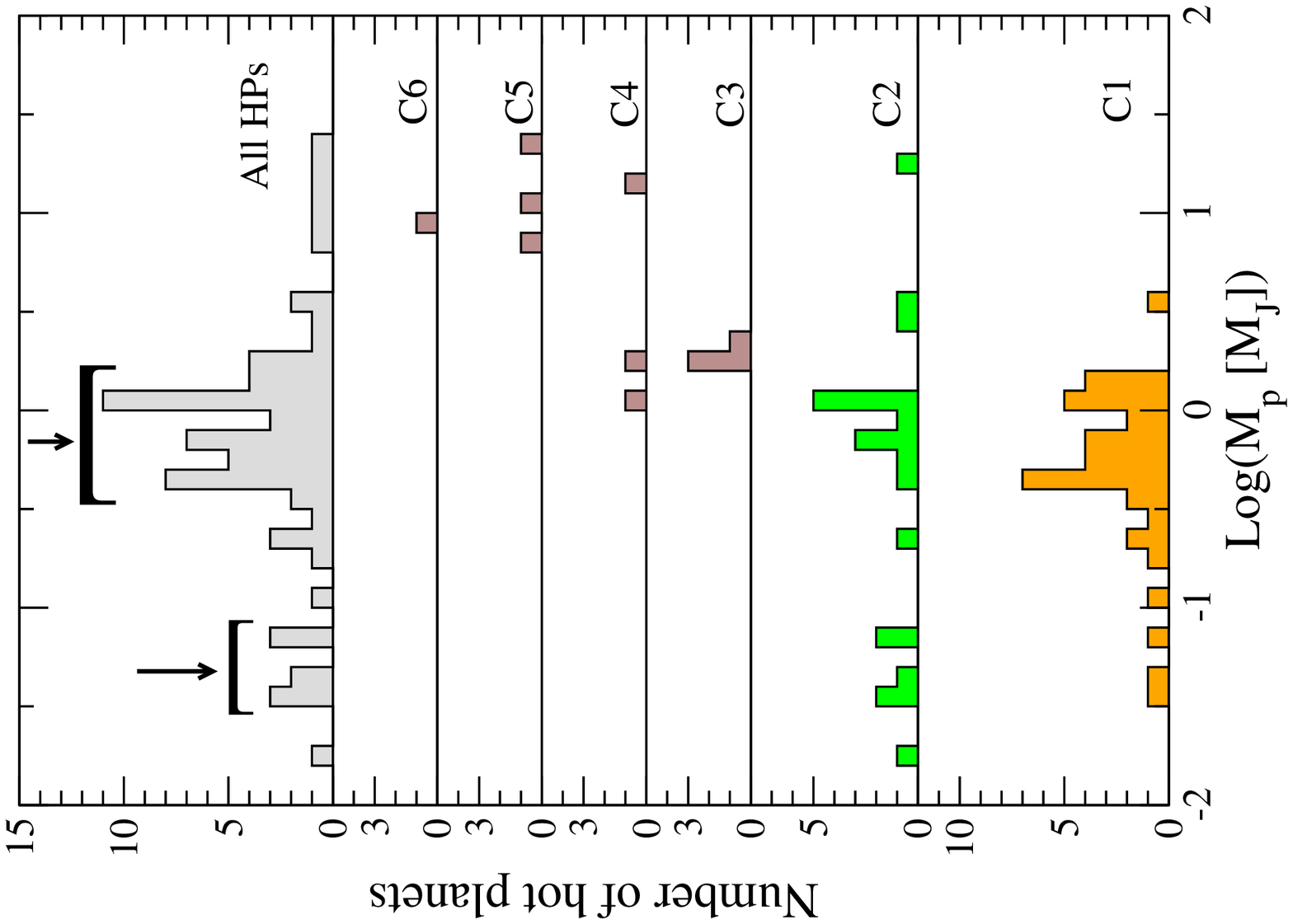}}
\subfigure{\includegraphics[width=0.4\textwidth,angle=-90]{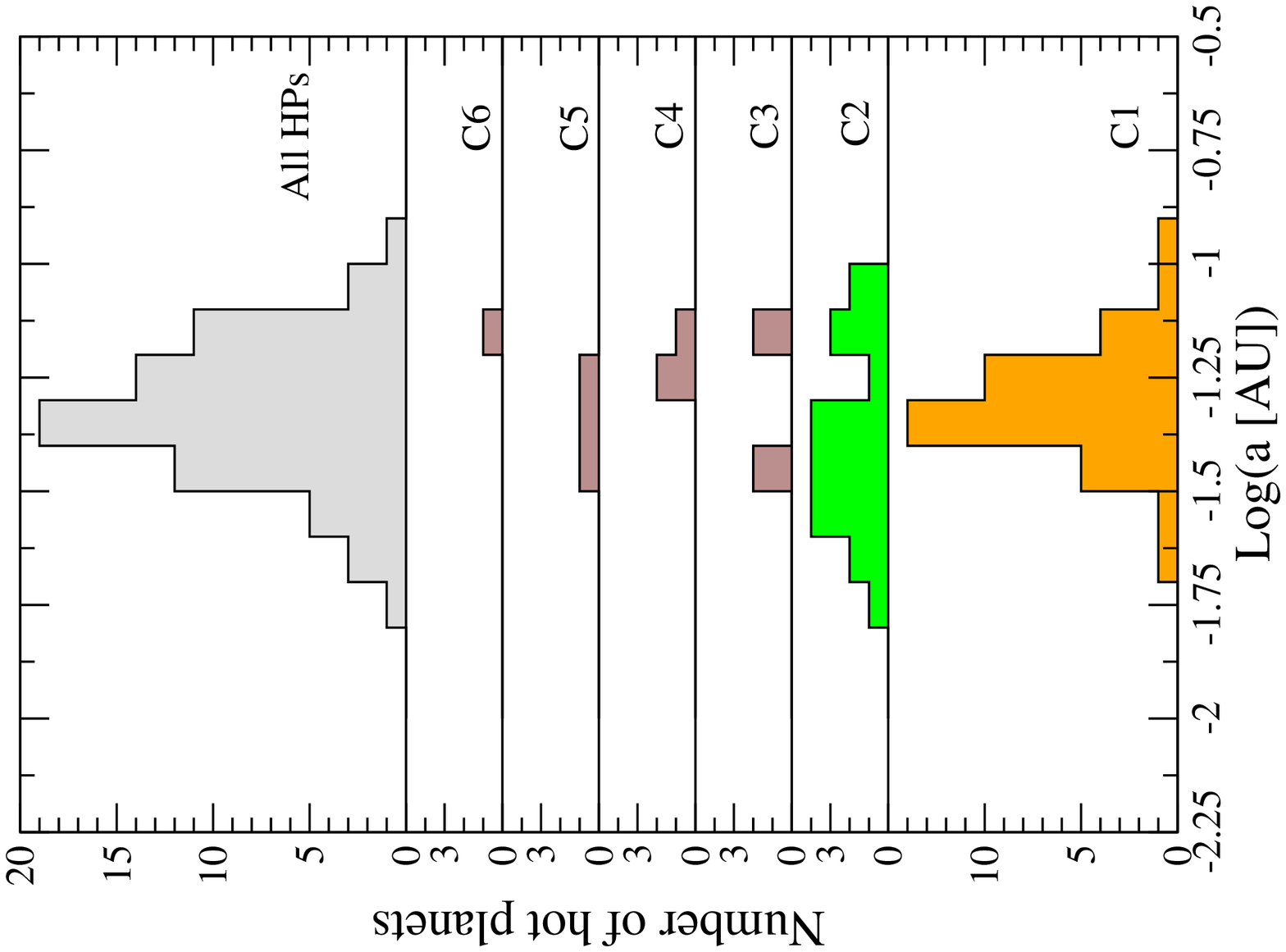}}
\caption{Left  panel:  Hot  planet  mass distributions.  The  current
  distribution of all HPs shows two peaks (indicated by arrows). Right
  panel: HP semimajor axis  distributions.  Hot planets of clusters C1
  and   C2   differ  considerably   in   terms   of  their semimajor   axis
  distributions.   HPs  of  C2  have    quite a flat  semimajor  axis
  distribution, while those of C1  seem to be narrowly peaked at 
  $log(a)\sim-1.35$ ($a\sim0.045$~AU).}
\label{mp_a_hist}
\end{figure*}

\begin{figure*}
\subfigure{\includegraphics[width=0.4\textwidth,angle=-90]{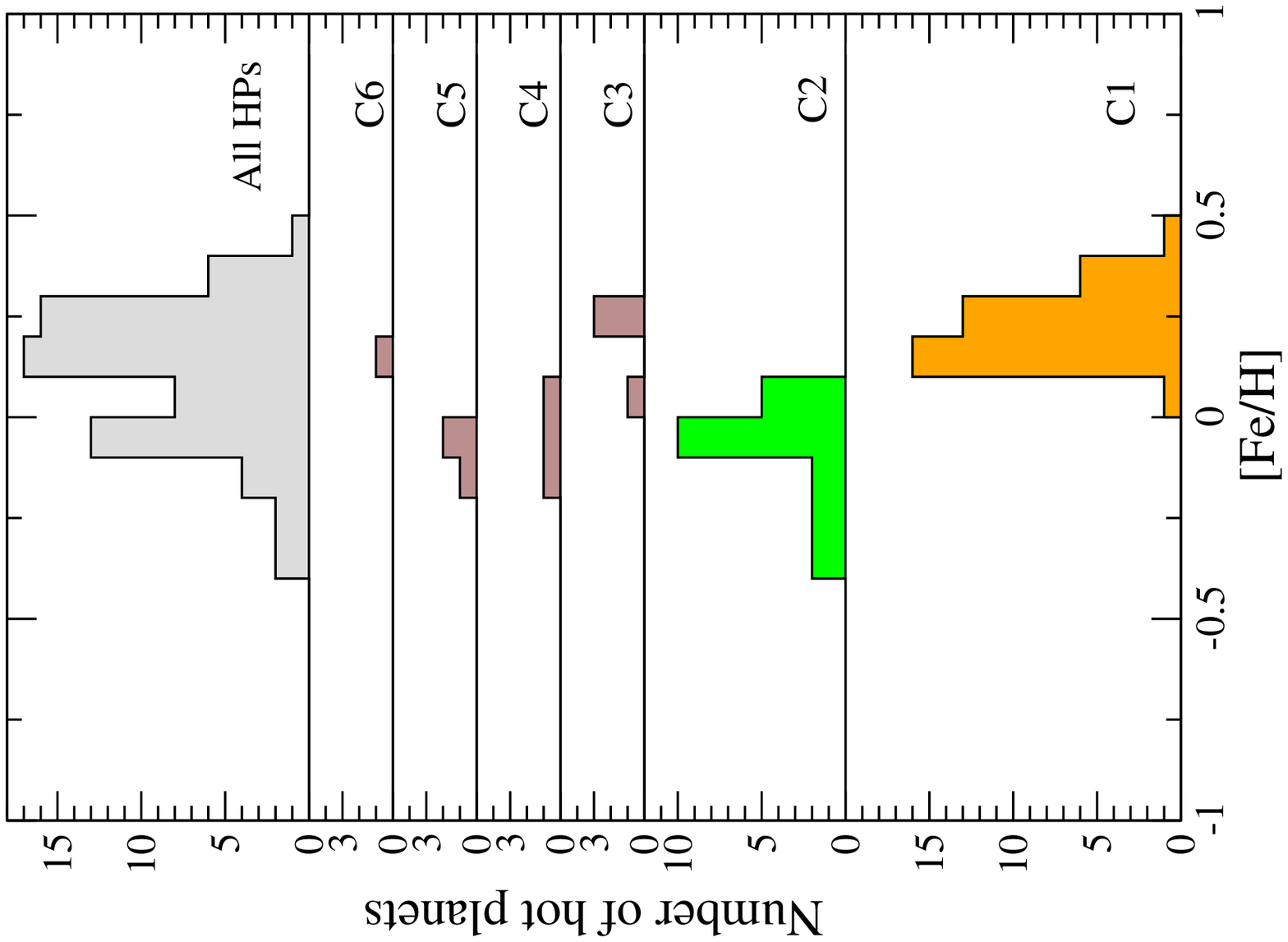}}
\subfigure{\includegraphics[width=0.4\textwidth,angle=-90]{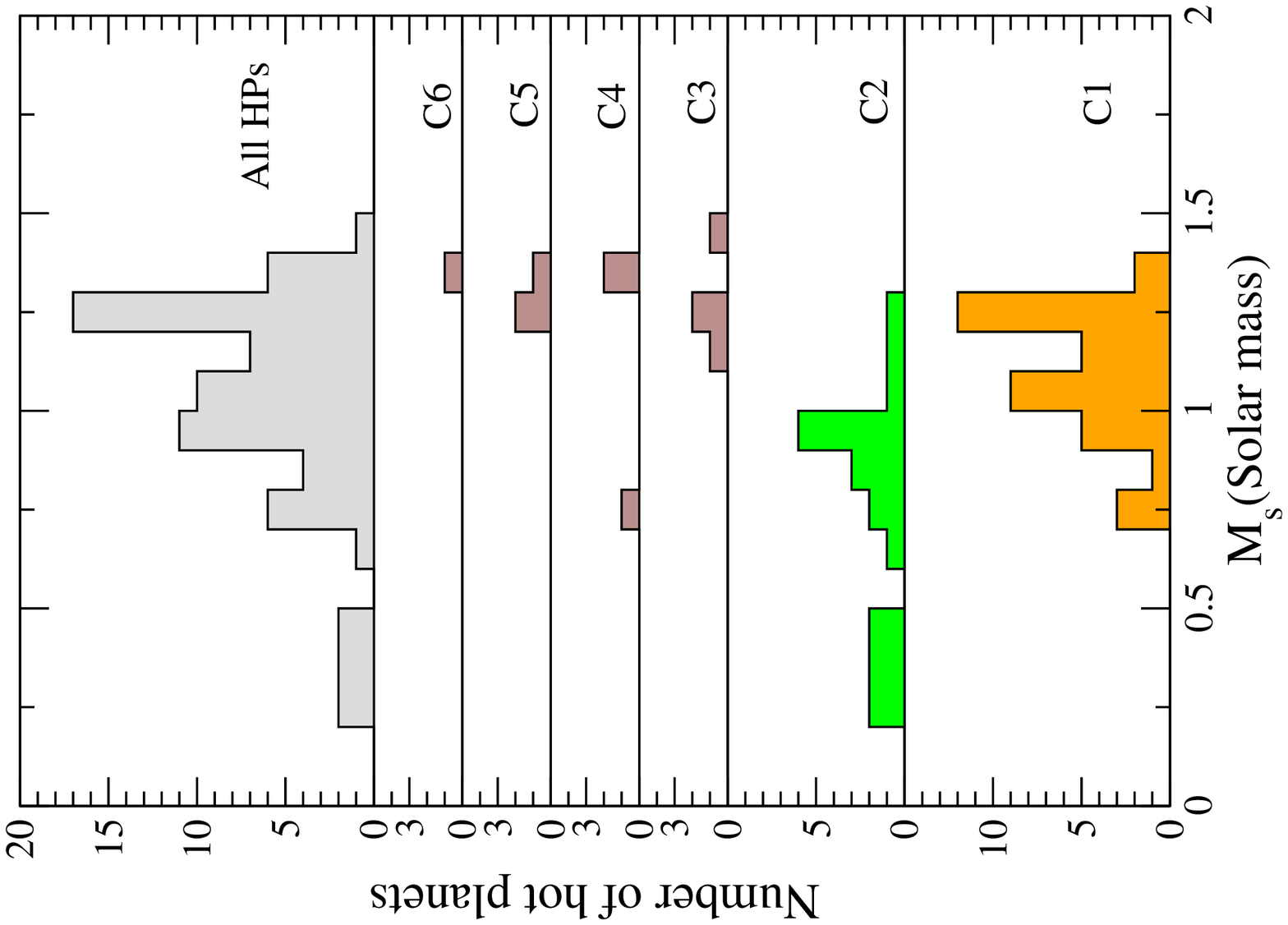}}
\caption{Left panel: HP metallicity  distributions.  HPs of cluster C1
have super-solar  [Fe/H], while those  of C2 have a  sub-solar [Fe/H].
Right  panel: HP  stellar mass  distributions. HPs  of C1  have mostly
super-solar $M_s$; while many HPs in C2 have sub-solar $M_s$.}
\label{fe_ms_hist}
\end{figure*}

\begin{figure*}
\subfigure{\includegraphics[width=0.4\textwidth,angle=-90]{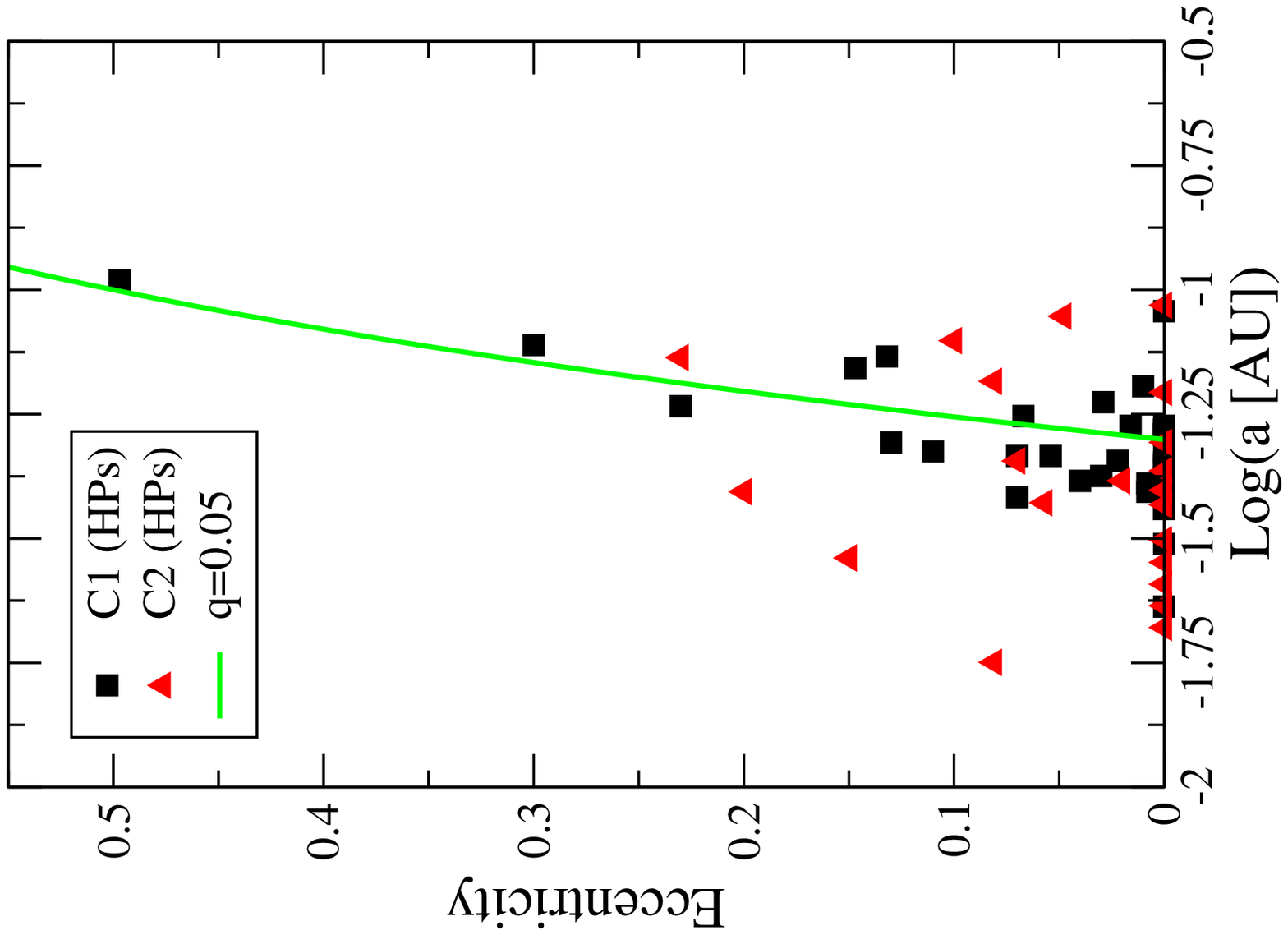}}
\hspace*{-4cm}
\subfigure{\includegraphics[width=0.4\textwidth, angle=-90]{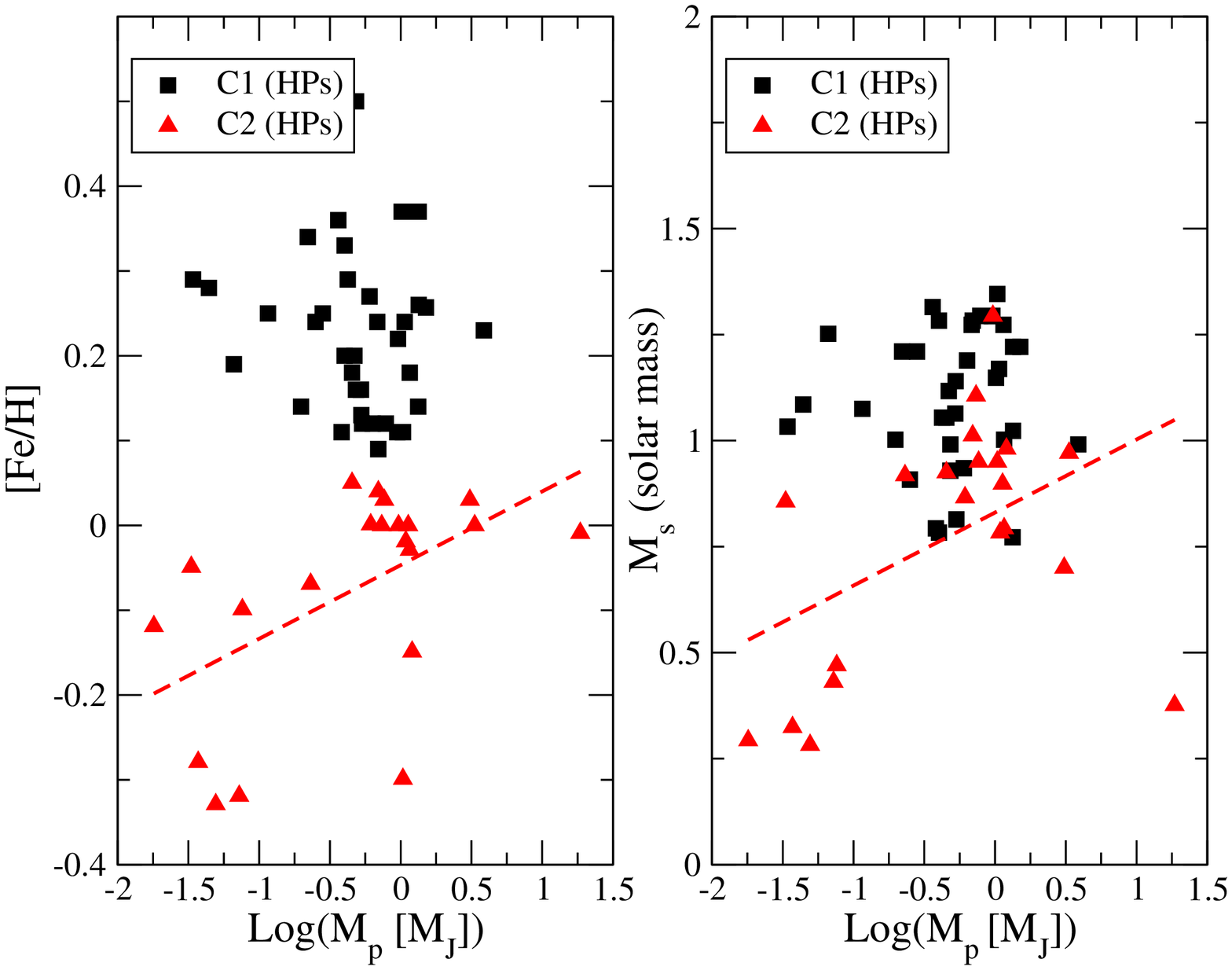}}
%\centerline{\psfig{figure=hj_ae_new.ps,width=11cm,angle=-90}}
\caption{Significant  intracluster  correlations.   Left panel:  $a-e$
distribution  for cluster  C1 and  C2. The  two groups  of HPs  show a
remarkable   difference:  C1  shows   a  very   strong,  statistically
significant, correlation between $a$  and $e$ (2-tailed probability of
0.01\%), while  C2 has no  correlation. Moreover, cluster  C2 exhibits
two statistically significant  correlations. They are the $M_p-$[Fe/H]
correlation  (middle panel) and  $M_p-M_s$ correlation  (right panel),
respectively  with  a  2-tailed  probability  of  1\%  and  4\%.  Such
correlations do not hold for cluster C1.}
\label{ae}
\end{figure*}

One of  the most interesting features  about the HPs, is  the shape of
their mass distribution (see fig.~\ref{mp_a_hist}).  It has two peaks.
These peaks are  thought to be real, even if the  shape of the present
distribution is  severely affected  by observational biases  that make
the  discovery of  low mass  HPs  difficult.  In  the literature,  the
bodies  belonging  to the  lower  mass peak  are  referred  to as  hot
neptunes (HNs)  while those belonging to  the higher mass  peak as hot
jupiters  (HJs).   Theoretical models  also  predict  a double  peaked
distribution \citep{mor07,ida08b},  where HNs are expected  to be much
more abundant than HJs.\\
It is  interesting that our taxonomy  splits the majority  of HPs into
two different  clusters (except for a  few outliers). Most  of the HPs
belonging  to the  peak at  $log(M_p)\sim0$ are  placed in  cluster C1
(fig.~\ref{mp_a_hist}).   HPs  of the  cluster  C2  have  a broad  and
flatter  mass   distribution,  and  it  contains  most   of  the  HNs.
Concerning  the semimajor  axis  distribution, similar  considerations
hold.  HPs  of cluster C1 are strongly  clustered at $log(a)\sim-1.35$
($a\sim0.045$~AU);  while  those of  cluster  C2  have  a flatter  $a$
distribution (fig.~\ref{mp_a_hist}).  The two groups of  HPs also have
clearly  distinctive   traits  in  terms  of   properties  of  stellar
metallicity and -to a lesser extent- to stellar mass.  These variables
are  important because  they  account for  the  environment where  the
planets  formed.  Figure~\ref{fe_ms_hist}  shows a  remarkable result:
HPs of  cluster C1 have super-solar  [Fe/H], while those of  C2 have a
sub-solar [Fe/H].  Notice that this result is also valid for the whole
of  C1 and  C2,  and not  only for  their  HPs.  A  similar, but  less
pronounced, result  also holds  for the stellar  mass: HPs of  C1 have
mostly  super-solar $M_s$;  while those  of C2  have  mostly sub-solar
$M_s$ (fig.~\ref{fe_ms_hist}).\\
The HPs  of cluster C1 and  C2 also have other  distinctive traits. We
found some  significant intracluster  correlations which hold  for one
cluster  but not  for the  other.  These  are the  correlations $a-e$,
$M_p-$[Fe/H],  $M_p-M_s$ (see fig.~\ref{ae}).   The semimajor  axis of
the  HPs  of  cluster  C1  strongly correlate  with  $e$,  while  this
correlation  is absent  for cluster  C2.  On  the other  hand,  HPs of
cluster C2 exhibit a strong correlation of $M_p-$[Fe/H] and $M_p-M_s$,
while cluster  C1 does not  (fig.~\ref{ae}).  These two plots  in turn
clearly show  that HNs of cluster  C1 are very different  from the few
belonging to cluster C2.   The latter have remarkable sub-solar [Fe/H]
and sub-solar $M_s$.\\
Another important point  is why some HPs have  been placed in clusters
other  than C1  and C2.   They are:  HD73256b, HD68988b,  HAT-P-7b and
HD118203b  (C3);   XO-4b,  HAT-P-6b  and   HD162020b  (C4);  WASP-14b,
CoRoT-Exo-3b and XO-3b  (C5); HAT-P-2b (C6).  It is  not clear if they
are  real  outliers  or  if  they  are  misplaced  by  the  clustering
algorithm. However,  a close look  at their properties shows  that the
main characteristic of these HPs is that they have high $M_p$ and high
$M_s$.  Among the outliers, the most massive ($M_p\sim10~M_J$) HPs are
present, namely HAT-P-2b,  WASP-14b, CoRoT-Exo-3b, XO-3b and HD162020b
(see  fig.~~\ref{mp_a_hist},  left panel).   Notice  that the  massive
HD41004Bb is placed  instead in C2, due to the  very low stellar mass.
In the present database, the  combination of high $M_p$ and high $M_s$
is  rather unusual,  and  this suggests  that  these HPs  may be  real
outliers.   Of course,  if more  HPs having  such  characteristics are
discovered, it is possible that they may be grouped into a third class
of HPs. 
%
%Finally,  stellar multiplicity  (MSS) my  play an  important  role for
%planetary     evolution,     and     also     for     HP     formation
%\citep[e.g.][]{wu07,mar00}.   The HPs  of  cluster C1  and  C2 have  a
%similar percentage of MSS. 
Finally, it is also important to note that previous results are robust
to changes in the selection of  HP periods, at least in the range from
10~d to 30~d.

\section{Physical interpretation}

\begin{figure*}
\includegraphics[width=0.4\textwidth,angle=-90]{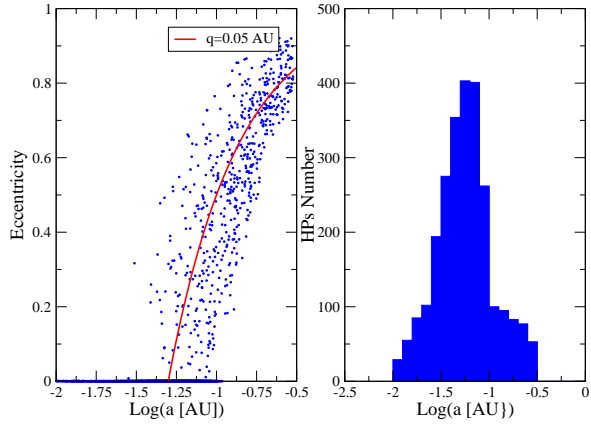}
%\centerline{\psfig{figure=scattering_ae_ahist.ps,width=11cm,angle=-90}}
\caption{Details of  the final  orbital configuration obtained  in the
 scattering  model.  As  for  the initial  conditions,  we adopted  an
 uniform  distribution   (in  linear  scale)  of   $M_p$,  an  uniform
 distribution in planetary radius,  and a gaussian $M_s$ distribution,
 peaked  at $1~M_{\odot}$.   Initial  semimajor axes  were choosen  at
 $\sim  1.35 M_s^2$~AU  (half  of the  snow-line  distance).  The  $e$
 distribution  has  been taken  according  to  fig.~9  of Nagasawa  et
 al.~2008, where  only $e>0.9$ is interesting for  the tidal evolution
 (see text for more details).}
\label{scatter_ae}
\end{figure*}

Our present theoretical understanding of the formation of HPs is based
on  two  models:  planetary  migration within  a  protoplanetary  disk
\citep{lin85}   and  planet-planet   scattering   followed  by   tidal
circularization \citep{ras96, wei96,  lin97}.  Planetary migration has
been shown to be an  efficient mechanism to produce HPs. Migration can
occur  when  planetary embryos  are  still  embedded  within the  disk
(type~I), or  when they  are large enough  to open  a gap in  the disk
(type~II). According  to the present  state of the art,  the migration
seems the best  candidate for HP formation.  On  the other hand, early
works on planet-planet scattering  showed that the probability for the
scattering model  to produce HPs was very  low.  However, \cite{nag08}
found that the Kozai mechanism enhances the probability significantly.
Therefore,  it is  possible  that the  scattering  contributes to  the
formation of  hot planets as  well as type~II migration,  although the
latter may be a main channel.\\
Type~II migration  is more efficient  for moderate mass  planets since
the planets have to be massive enough  to clear a gap in the disk, but
not too  massive to efficiently  exchange angular momentum  and energy
with the disk itself.
%On the other  hand, the scattering is more efficient for
%large mass  planets because  tidal deformation by  host stars  is more
%efficient  for  larger  planets.    
Previous  simulations  have  shown  that  the  formation  of  HPs  via
gravitational  scattering  among  planets  and  the  subsequent  Kozai
mechanism  combined   with  tidal  dissipation  is   more  likely  for
dynamically active  systems of multiple  planets, typically containing
three or more gas giants \citep[e.g.][]{mar02}.  The formation of many
giant  planets  is  preferred  in  high  dust  surface  density  disks
\citep{ida08a}.  Dust surface  density scales as $10^{\rm [Fe/H]}M_d$,
where $M_d$  is the total disk  mass. The latter  scales, according to
theoretical models, as  $\sim M_s^k$, where $0.7<k<2.2$ \citep{vor08}.
Therefore high  stellar masses  and high-metallicity disks  favor the
onset of  a scattering phase.   It must  also be noted  that smaller
planets tend  to be scattered  inward during the scattering  phase and
that  the tidal  circularization is  more efficient  for  planets with
small  mass and  large physical  radius.  
%It must also  be noted that during  the scattering
%phase, 
%the lower  mass planets present in a  multiple planetary system
%are those  more efficietly pushed inward.  
The detailed final orbital configuration of the HPs may vary according
to  several  parameters.   However,  there  are a  number  of  general
features that can be outlined.\\
 In the case of type~II migration,  the HPs final position is close to
the  location of  the  disk's inner  edge.   This is  placed near  the
corotation  radius (namely  the  distance where  the keplerian  period
matchs the  star spin period),  where disk material accretes  onto the
stellar poles following the magnetic field lines.  The spin period for
young   stars   having   $0.4~M_{\odot}<M_s<1.5~M_{\odot}$  may   vary
considerably, from  1~d up to 20~d or  more \citep{her07}.  Therefore,
taking also into account a  wide variety of disk parameters, the final
location of HPs formed by type~II migration tends to be spread, in the
range from 0.01 to 0.1~AU.  Concerning the eccentricity, at the end of
the migration  phase low $e$ values are  expected.  Recent simulations
\citep{ric08} suggest  that, when HPs reach  the stellar magnetosphere
cavities,   they   may   further   evolve   to   smaller   $a$.    For
moderate-to-high $M_p$,  the $e$ also  increases.  On the  other hand,
the location of HPs formed by scattering is determined by the location
where  the tidal  force is  effective. The  tidal strength  depends on
several  parameters (planetary  radius, planetary  mass, etc)  but the
final  location tends  to be  in the  range 0.03-0.08~AU,  for typical
parameter values.  The eccentricity of  the inner planet is excited to
values close to unity by close scattering and via the subsequent Kozai
mechanism.   The  resultant  small  pericenter  distance  enables  the
planet's eccentricity and semi-major axis to be decreased by the tidal
dissipation  and  moderate eccentricities  can  remain  in some  cases
\citep{nag08}.  We show this in fig.~\ref{scatter_ae}.  The outcome of
the  simulations   vary  according   to  tide  efficiency.    For  the
simulations  shown  here, we  followed  the  tidal  evolution of  test
planets,   according   to   \cite{iva07},   for   a   time   span   of
$10^9-10^{10}$~yr.   A  large  number  of planetary  radii,  planetary
masses  and stellar  masses have  been considered,  choosen  at random
within     the     following     intervals:     $0.4~R_J<R_p<2.8~R_J$,
$0.5~M_J<M_p<3~M_J$,  $0.5~M_{\odot}<M_s<1.5~M_{\odot}$.   The initial
eccentricity  is randomly  chosen  from the  distribution obtained  by
\cite{nag08}.  Figure~\ref{scatter_ae} (left  panel) shows that at the
end of the scattering and  tidal evolution phases the eccentricity and
semimajor axes  are correlated (compared to  the observed distribution
in  fig.~\ref{ae}), while  the right  panel shows  that  the resulting
semimajor axis distribution is peaked.   The value of the peak depends
on the strength of the tide, and for the values used here it is peaked
at  $a\sim 0.05$~AU,  as  observed  in cluster  C1  (compare with  the
observed distribution in fig.~\ref{mp_a_hist}).

\section{Discussion and conclusion}

The  two main  processes  of HP  formation  produce different  orbital
distributions. On the basis of our taxonomic analysis, we identify two
types of HPs,  which may retain the footprints  of these two different
formation processes.   In this respect, HPs  of cluster C2  and C1 may
have  been formed  by  migration and  scattering, respectively.   This
scenario  is supported  by a  number of  facts. First  of all  the $a$
distribution and  the strong $a-e$ relationship for  cluster C1, which
do not hold  for cluster C2. Moreover, also  the $M_p$ distribution of
the  two clusters  support  this conclusion:  higher  mass bodies  are
located in  cluster C1, indicating more  massive protoplanetary disks,
which in  turn is confirmed  by high [Fe/H]  and high $M_s$.   In this
case, the detected  HPs are expected to be the  least massive for each
system, since in a multibody  scattering the least massive planets are
more  effectively  pushed inward.   The  orbits  of  the most  massive
planets are, however, only  slightly affected by the scattering phase,
therefore  they tend  to stay  close  to their  formation regions  and
therefore on  relatively large $a$.   Although HJs originated  via the
scattering process are expected to be accompanied by outer companions,
the latter would be beyond the detectable limit by present surveys. We
find  no   significant  difference  between   C1  and  C2   about  the
multiplicity of planets, but we  caution that this result is affected
by low number statistics.  Some Jupiter-like mass HPs are also present
in cluster  C2, but in this case  they may be the  most massive bodies
formed in these systems, given  also the moderate-to-low $M_s$ and low
[Fe/H].  Another interesting  point is that most of  the HNs belong to
cluster C2.\\
Therefore, if we extrapolate  directly the percentage of HPs belonging
to C1 and C2 into the efficiency of formation of the two processes, we
end up with 50\% of HPs  formed via scattering and 30\% via migration.
The remaining 20\%  are outliers and may be  formed either way.  These
numbers, however, have to be  taken with caution, since some degree of
mixing between the two clusters is expected. On the other hand, from a
theoretical  point  of  view  (see discussion  in  previous  sections)
type~II is  expected to be  more efficient in producing  close-in EPs.
This  fact is  not  in contradiction  with  our findings  since it  is
possible  that planets  migrating inward  by type~II  may  stop before
becoming HPs. This would be the  case, for instance, if the gas in the
disk   dissipates  before  the   planet  reaches   the  magnetospheric
cavity. In this  respect, it is interesting that  many giant EPs exist
in the  range 0.1-3~AU and  that their semimajor axis  distribution is
well explained by the type~II migration model \citep{sch09}.\\
An alternative scenario  is that the two groups of  HPs were formed by
the same process, and the cluster analysis splits them on the basis of
their  diversity in  the  input  variables.  In  this  case, the  most
distinctive  variable  would be  the  metallicity.   Formation in  low
metallicity, moderate-to-low  $M_s$, environments would  have produced
the  HPs  of  cluster  C2.    On  the  other  hand,  high  [Fe/H]  and
moderate-to-high $M_s$ would have produced the HPs of cluster C1.  The
combinations  of  these  diversities  would  have  also  produced  the
observed  differences in  the $M_p$  and $a$  distribution of  the two
groups. Although this  is a possibility, we think  the peculiar traits
of cluster C1 and C2 clearly shown the influences of the two formation
mechanisms.
%In
%this  scenario, the  outliers may  have been  produced by  a different
%process and/or  by peculiar values  of the input variables  (e.g. high
%$M_s$, see fig. \ref{ms_hist}).\\

\bsp

\label{lastpage}


\begin{thebibliography}{99}

\bibitem[Herbst  et al.(2007)]{her07}  Herbst, W.,  Eisl{\"o}ffel, J.,
Mundt, R., \& Scholz, A.\ 2007, Protostars and Planets V, 297

\bibitem[Ida \&  Lin(2008)]{ida08a} Ida,  S., \& Lin,  D.~N.~C.\ 2008,
ApJ, 673, 487

\bibitem[Ida  \& Lin(2008)]{ida08b}  Ida, S.,  \& Lin,  D.~N.~C.\ 2008,
ApJ, 685, 584

\bibitem[Ivanov   \&   Papaloizou(2007)]{iva07}   Ivanov,  P.~B.,   \&
Papaloizou, J.~C.~B.\ 2007, MNRAS, 376, 682

\bibitem[Lin \& Ida(1997)]{lin97} Lin, D.~N.~C., \& Ida, S.\ 1997,
ApJ, 477, 781

\bibitem[Lin \& Papaloizou(1985)]{lin85} Lin, D.~N.~C., \& Papaloizou,
J.\ 1985, Protostars and Planets II, 981

\bibitem[Marchi \& Ortolani(2008)]{mar08} Marchi, S., \&
Ortolani, S.\ 2008, IAU Symposium, 249, 123

\bibitem[Marchi(2007)]{mar07} Marchi, S.\ 2007, ApJ, 666, 
475 

\bibitem[Marzari  \&  Weidenschilling(2002)]{mar02}  Marzari,  F.,  \&
Weidenschilling, S.~J.\ 2002, Icarus, 156, 570

%\bibitem[Marzari 
%\& Scholl(2000)]{mar00} Marzari, F., \& Scholl, H.\ 2000, ApJ, 543, 328 

\bibitem[Mordasini et al.(2007)]{mor07} Mordasini, C., 
Alibert, Y., Benz, W., \& Naef, D.\ 2007, arXiv:0710.5667 

\bibitem[Nagasawa et al.(2008)]{nag08} Nagasawa, M., Ida, S., 
\& Bessho, T.\ 2008, ApJ, 678, 498 

\bibitem[Rasio  \& Ford(1996)]{ras96}  Rasio, F.~A.,  \&  Ford, E.~B.\
1996, Science, 274, 954

\bibitem[Rice et al.(2008)]{ric08} Rice, W.~K.~M., Armitage, P.~J., \&
Hogg, D.~F.\ 2008, MNRAS, 384, 1242

\bibitem[Schlaufman et al.(2009)]{sch09} Schlaufman, K.~C., 
Lin, D.~N.~C., \& Ida, S.\ 2009, ApJ in press

\bibitem[Vorobyov(2008)]{vor08} Vorobyov, E.~I.\ 2008, 
arXiv:0810.1393 

\bibitem[Weidenschilling   \&  Marzari(1996)]{wei96}  Weidenschilling,
  S.~J., \& Marzari, F.\ 1996, Nature, 384, 619

%\bibitem[Wu et al.(2007)]{wu07} Wu, Y., Murray, N.~W., 
%\& Ramsahai, J.~M.\ 2007, ApJ, 670, 820 


\end{thebibliography}
\end{document}